\documentclass{ws-procs9x6}

\newcommand\Ref[1]     {Ref.\,\cite{#1}}
\newcommand\Refs[1]    {Refs.\,\cite{#1}}

\newcommand\Fig[1]     {Fig.\,{\ref{#1}}}

\newcommand\nn         {\nonumber}

\def\beq{\begin{equation}}
\def\eeq{\end{equation}}
\def\beeq{\begin{eqnarray}}
\def\eeeq{\end{eqnarray}}

\newcommand\as         {\ensuremath{\alpha_{\mathrm{s}}}}

\newcommand\msbar      {\ensuremath{{\overline {\rm MS}}}}
\newcommand\muR[1]     {\ensuremath{\mu_R^{#1}}}
\newcommand\muF[1]     {\ensuremath{\mu_F^{#1}}}

\newcommand\rNLO       {\mathrm{NLO}}

\newcommand{\rd}       {{\mathrm{d}}}

\newcommand\kT         {k_\perp}

\newcommand\pdf{parton distribution function}
\newcommand\LO{LO}
\newcommand\NLO{NLO}

\newcommand\xB    {\ensuremath{x_{\rm B}}}
\newcommand\ycut  {\ensuremath{y_{\rm cut}}}

\newcommand\Kout  {\ensuremath{K_{\rm out}}}
\newcommand\Et    {\ensuremath{E_{\rm t}}}

\newcommand\Qhs   {\ensuremath{Q_{\rm H.S.}}}

\begin{document}

\title{Multi-jet production in lepton-proton scattering at
next-to-leading order accuracy%
\footnote{\uppercase{c}ontribution to the \uppercase{p}roceedings of
the \uppercase{r}ingberg workshop on ``\uppercase{n}ew
\uppercase{t}rends in \uppercase{HERA} \uppercase{p}hysics 2005''.}}

\author{Z. TR\'OCS\'ANYI}

\address{University of Debrecen
and Institute of Nuclear Research of the Hungarian Academy of Science, \\
H-4001 Debrecen P.O.Box 51, Hungary, \\ 
E-mail: Z.Trocsanyi@atomki.hu}

\maketitle

\abstracts{
I summarize the theoretical and experimental status of multijet
production in DIS. I present the state of the art theoretical
predictions and compare those to the corresponding experimental results
obtained by analysing the data collected by the H1 and ZEUS
collaborations at HERA. I also show new predictions for three-jet
event-shape distributions at the NLO accuracy.
}

\vskip -10cm
\title{\hfill hep-ph/0512004}

\vskip 9cm

\section{Introduction}

Deep inelastic lepton-hadron scattering (DIS) has played a decisive role
in our understanding of the deep structure of matter. The latest version
of the experiment performed with colliding electrons or positrons and
protons at HERA yields increasingly precise data so that not only fully
inclusive measurements can be used to study the physics of hadronic
final states. In fact, the study of jet-rates and event shapes has
become an important project at HERA which yields results with
continuosly increasing accuracy \cite{HERA}.  Thus HERA is considered a
machine for performing precision measurments for understanding Quantum
Chromodynamics (QCD), the theory of strong interactions.
 
In order to perform precision measurements one needs precision tools for
analysing the data. In the case of studying hadronic final states in
high-energy particle collisions such tools have been developed in the
framework of perturbative QCD. In order to make precision quantitative
predictions in perturbative QCD, it is essential to perform the
computations (at least) at the next-to-leading order (NLO)
accuracy. Such computations however, yield reliable predictions only in a
limited part of the phase space, where the statistics of the data are
relatively small. In order to increase the predictive power of the
theory, the fixed-order predictions must be improved by matching those to
predictions obtained by resumming large logarithmic contributions to all
orders.

In the case of DIS fixed-order and resummation computations have so far
been completed for one-jet inclusive, 2- or 3 (+1 beam%
\footnote{In the following, I omit the reference to the beam-jet, i.e.,
I do not count the beam-jet.}%
)-jet cross sections and means or distributions of event shapes. Since
the main goal of the experimental analyses is to compare data to
precision theoretical predictions (beyond the LO accuracy), this
implies that the experimental analysis of multi-jet events is generally
constrained to considering three-jet events.  Therefore, in this talk
``multi'' will mean three.  This is in contrast to hadron collider studies,
where the multi-jet events are backgrounds to various new-particle
signatures, therefore, even the predictions at LO are considered
valuable information and the construction of parton-level event
generators is an important research topic \cite{Mangano:2003ps}. One of
the main lines of this research is the construction of public computer
programs that could be used for the automated production of
multi-parton events and thus, for computing multi-jet cross sections.
These computer programs could also be used to study high-multiplicity
final states in DIS. Note however, that the studies that can be made at
HERA are not directly applicable at the LHC because the events at HERA
has jets of typical energy in the order of 10\,GeV while jets at the
LHC will be triggered at the order of 100\,GeV.

The automation of computing cross sections at the NLO accuracy has also
been considered, but has not yet yielded mature results. Thus one has to
recourse to programs for specific processes.  The state of the art in
the fixed-order computations of cross sections in DIS is represented by
the {\sc nlojet++} program that can be used for computing two- and
three-jet observables \cite{NTdisPRL,nlojet}. Other related programs for
leptoproduction of two jets are {\sc disent}, {\sc disaster++} and {\sc
jetvip} \cite{disent,disaster,jetvip}. The predictions obtained by
the {\sc nlojet++} and {\sc disaster++} codes for two-jet
cross sections agree within statistical accuracy of the numerical
integrations.\footnote{The essential difference is that {\sc nlojet++}
is significantly faster.} The {\sc disent} code is known to have a
small bug \cite{DSdisresum} leading to slightly different predictions
(the cross sections agree within 1--2\,\% \cite{disaster}).
At the time of the comparison of {\sc disent} and {\sc jetvip} the
latter code was not able to produce reliable predictions over the whole
phase space \cite{Duprel:1999wz}, which was due to a bug in the binning
routine that has been corrected since \cite{klasenprivate}.

The state of the art in the resummed predictions is represented by the
recent analytic computations of the distribution of the multijet event
shape \Kout\ \cite{Kout}, the di-jet rates with symmetric \Et\ cuts
\cite{dijet}, as well as by the {\sc caesar} program that can be used
for computing cross sections of two- and three-jet event shapes in a
semi-automatic way \cite{caesar}.  In my talk I shall present NLO
predictions of three-jet event-shape distributions for which resummed
predictions already exist, but the fixed-order radiative corrections
have not been computed before.  

\section{Fixed-order predictions}

There are several process-independent ways to compute QCD radiative
corrections.  In computing the \NLO\ corrections to multijet cross
sections, the dipole subtraction scheme of Catani and Seymour
\cite{CSdipole} is a convenient formalism. It is used both in the
{\sc disent} and the {\sc nlojet++} programs.
The comparison of the two-jet predictions of the three programs has
been performed in \Ref{NTdisPRL} and complete agreement was found apart
from the slight difference in the {\sc disent} predictions mentioned
above.  These programs use matrix elements that take into account only
virtual photon exchange.  Neglecting the exchange of the $Z^0$ boson
means that the predictions are not reliable for large $Q^2$ values
around (90\,GeV)$^2$ and above.
 
The subtraction scheme applied in the {\sc nlojet++} program is modified
slightly as compared to the original one in \cite{CSdipole} in order to
have a better control on the numerical computation. The main idea is to
cut the phase space of the dipole subtraction terms as introduced in
\Ref{NT4jets}. The details of the computations are given in
\Ref{NTdisevshape}.

Once the phase space integrations are carried out, one can write the NLO
jet cross section in the following form:
\beeq
&&
\sigma^{(J)}(p, q) =
\\ \nn &&\qquad 
\sum_{a} \int_0^1 \rd\eta
\,f_{a/P}(\eta, \mu_F^2)\,\sigma_{a,\rNLO}^{(J)}
\left(\eta p, q,\as(\mu_R^2), \muR 2/Q_{HS}^2, \muF 2/Q_{HS}^2\right)\:,
\label{signjet}
\eeeq
where $p^\mu$ and $q^\mu$ are the four-momenta of the incoming proton and
the exchanged virtual photon, respectively. The function
$f_{a/P}(\eta,\mu_F^2)$ is the density of the parton of type $a$ in the
incoming proton at momentum fraction $\eta$ and factorization scale
$\mu_F$.  The partonic cross section $\sigma_{a,\rNLO}^{(J)}$
represents the sum of the LO and NLO contributions, given explicitly in
\Ref{NTdisevshape}, with jet function $J$.  In addition to the parton
momenta and possible parameters of the jet function, it also depends
explicitly on the renormalized strong coupling $\as(\muR 2)$, the
renormalization and factorization scales $\muR{} = x_R \Qhs$ and
$\muF{} = x_F \Qhs$, where $\Qhs$ is the hard scale that characterizes
the parton scattering, set event by event. Furthermore, the cross
section also depends on the electromagnetic coupling, for which the {\sc
nlojet++} code uses $\msbar$ running $\alpha_{\scriptscriptstyle \rm
EM}(Q^2)$ at the scale of the virtual photon momentum squared $Q^2=-q^2$.

The publicly available version of the {\sc nlojet++} program
\cite{nlojet} is based on the tree-level and one-loop matrix elements
given in \Refs{NT4jets,BDK}, crossed into the photon-parton channel. It
uses a C/C++ implementation  of the LHAPDF library
\cite{Giele:2001mr} with CTEQ6M \cite{Pumplin:2002vw} parton
distribution functions and with the corresponding $\as$ expression
for the renormalized coupling which is included in this library.  The
CTEQ6M set was fitted using the two-loop running coupling with
$\as(M_{Z^0}) = 0.118$.

\section{Comparison of fixed-order predictions to data}

\begin{figure}[ht]
\centerline{
\epsfxsize=55mm \epsfbox{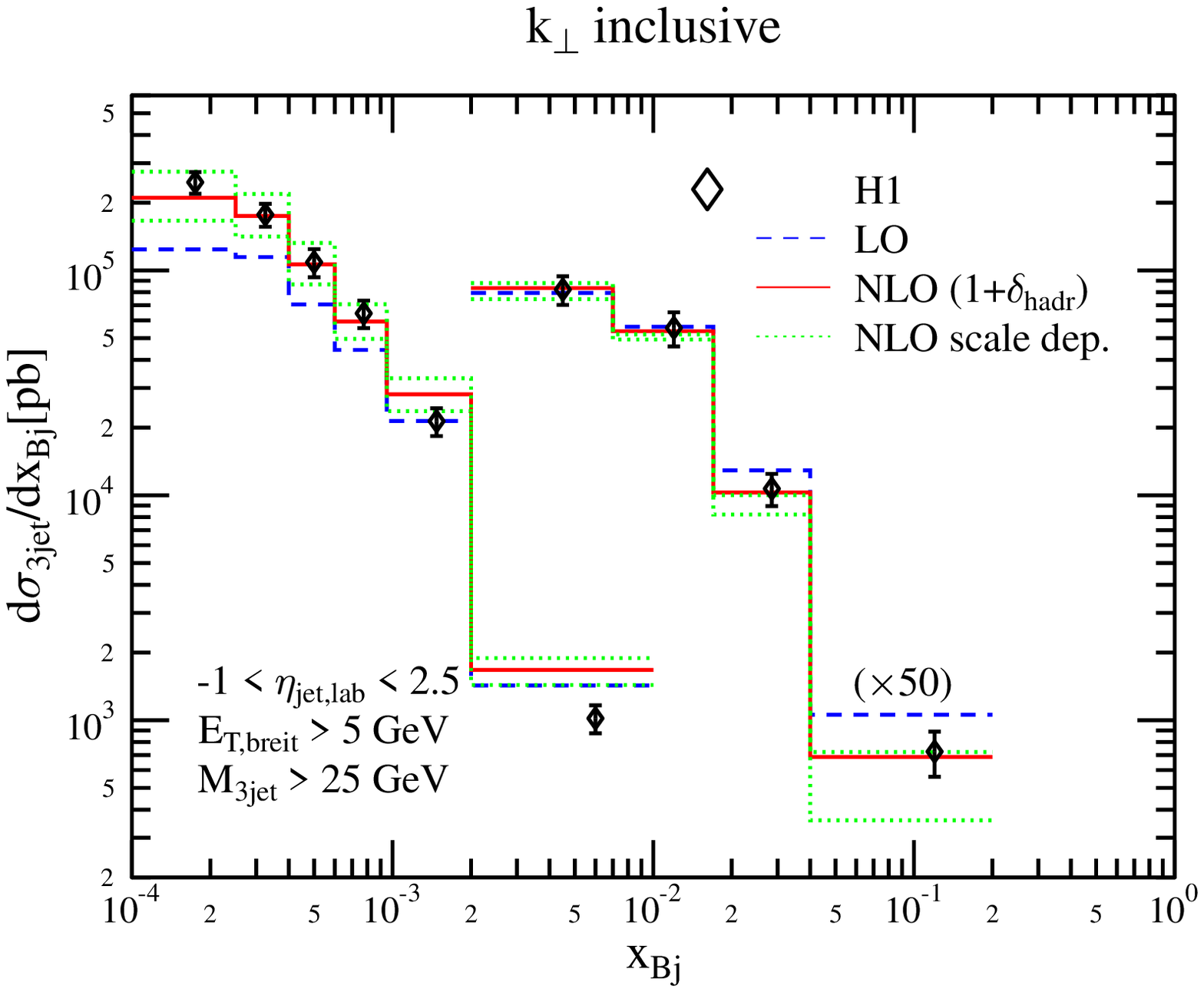}
~
\epsfxsize=55mm \epsfbox{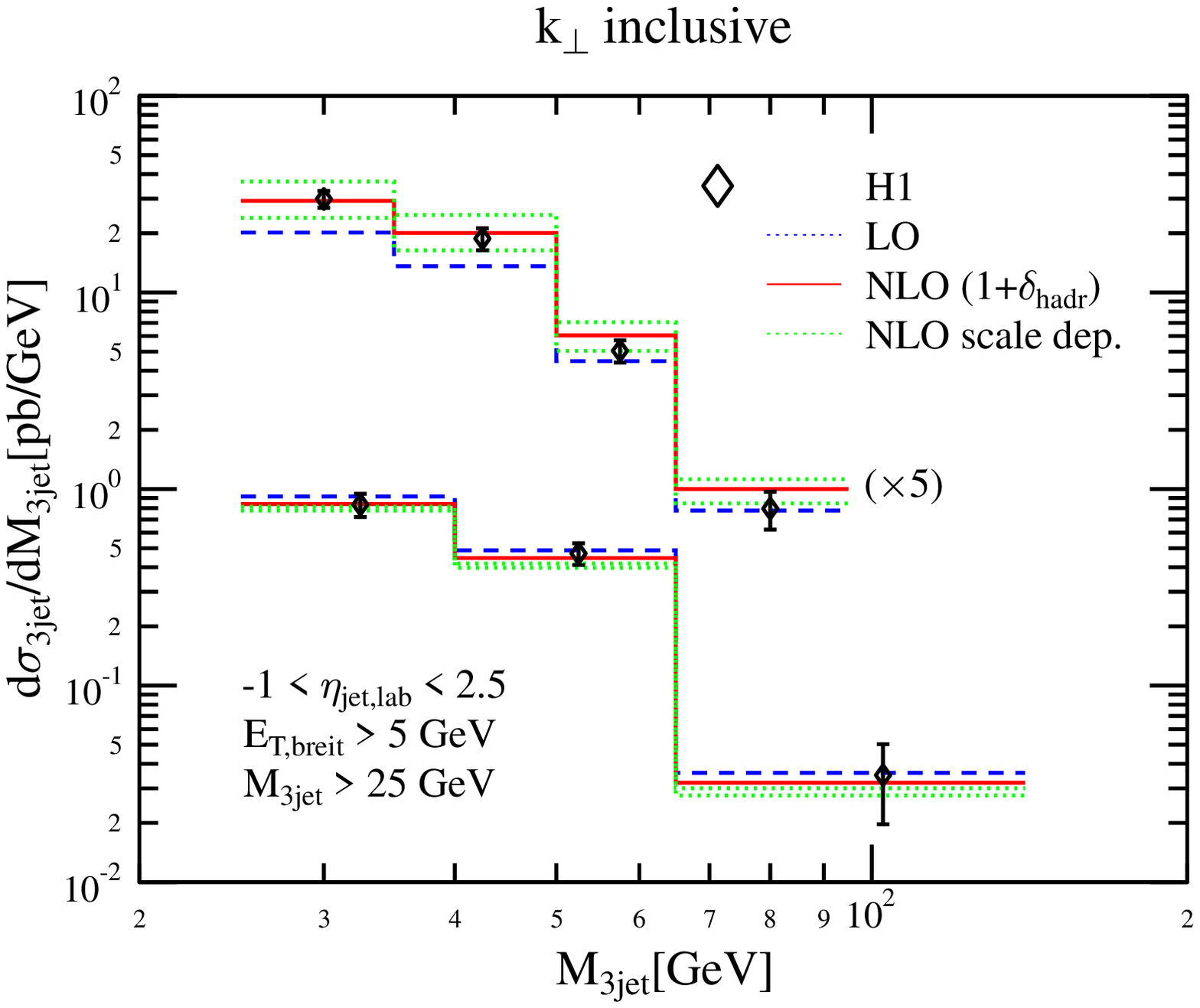}
}
\caption{The differential distributions of the Bjorken variable $\xB$
and the three-jet invariant mass $M_{\rm 3jet}$. The histograms with
larger values of the cross section correspond to the range 
$5\,{\rm GeV}^2 < Q^2 < 100\,{\rm GeV}^2$
and those with lower values to 
$150\,{\rm GeV}^2 < Q^2 < 5000\,{\rm GeV}^2$
\label{fig:H13jet}}
\end{figure}

During the last few years, the experimental groups at HERA has
performed extensive studies of multijet cross sections and compared
their results to NLO predictions. The H1 collaboration already
presented their results at this workshop four years ago
\cite{Wing:2001bt}. The analysis was carried out parallel to our
theoretical work with Z. Nagy that lead to the {\sc nlojet++} code, but
without knowing about each other. When we finished testing our program
and started to think of what to compute, we learnt about the H1 analysis
accidentally. At that time preliminary H1 results showed rather large
differences between data and LO predictions as seen on
\Fig{fig:H13jet}, even in the shapes of distributions, not only the
absolute normalization.%
\footnote{For obtaining the predictions at LO accuracy we used the
CTEQ5L \pdf s \cite{CTEQ5} and the running coupling at one-loop with
$\as(M_Z) = 0.127$.}
We decided to make predictions of the same distributions at NLO accuracy.

H1 defined the jets using the inclusive $\kT$ algorithm implemented in
the Breit frame (the precise definition can be found in
Ref.~\cite{Adloff:1999ni}), selected three-jet events and plotted
differential distributions of the DIS kinematical variables $Q^2$,
$\xB$ and the invariant three-jet mass $M_{\rm 3jet}$. We used the
same jet algorithm. Furthermore, in our computations we chose the same
kinematical region as H1 did \cite{Adloff:2001kg}, namely, for the
basic DIS kinematic variables $Q^2$, $\xB$ and $y = Q^2/(s\,\xB)$ we
required
$
5\,{\rm GeV}^2 < Q^2 < 5000\,{\rm GeV}^2\,,\;
0 < x_{\rm Bj} < 1\,,\;
0.2 < y < 0.6\:
$.
Following the H1 analysis, we also restricted the
(pseudo)rapidity-range in the laboratory frame and the minimum
transverse energy of the jets in the Breit frame as
$
-1 < \eta_{\rm lab}^{\rm jet} < 2.5\,,\;
E_{\rm T,B}^{\rm jet} > 5\,{\rm GeV}\:
$.
For the hard scattering scale we chose the average transverse momentum
of the jets, 
\beq
\label{eq:Qhs}
\Qhs = \frac13 \sum_j E_{\rm T,B}^{{\rm jet},\,j}\:.
\eeq
We also studied the other usual choice, when $\Qhs^2 = Q^2$, but have
not found significant differences. Finally, in order to compare our
parton-level prediction to the hadron level data, we asked for the
bin-wise correction factors of hadronization as estimated by H1 
(the correction factors were between 1.2--1.3). With the inclusion of the
NLO corrections the improvement in the theoretical description was
spectacular, see \Fig{fig:H13jet}.

Recently, the ZEUS collaboration has also performed an analysis of the
three-jet events \cite{ZEUS3jet}. They published measurements of the
inclusive three-jet cross section as a function of $Q^2$, the jet
transverse energy in the Breit frame, $E_{\rm T, B}^{\rm jet}$ and the
jet pseudorapidity in the laboratory frame $\eta_{\rm lab}^{\rm jet}$
compared to the NLO predictions obtained with the {\sc nlojet++} code,
corrected for hadronization (the correction factors $C_{\rm had}$ were
in the range of 1.15--1.35.)
The NLO QCD predictions were found to describe both the shapes of the
predictions as well as the absolute normalization of the measured cross
sections. The two competing most significant sources of uncertainty are
the energy scale uncertainty from the experimental side and
renormalization scale uncertainty from the theoretical side.

Such an agreement between data and theory promised a precise
measurement of the strong coupling and its running by fitting the cross
section ratio $R_{3/2}$ of the three-jet cross section to the two-jet one 
as a function of $Q^2$. The correlated systematic and
renormalization-scale uncertainties mostly cancel in the ratio. According
to the studies made by ZEUS \cite{ZEUS3jet}, the total experimental and
theoretical uncertainties are about 5\,\% and 7\,\%, respectively. The
reduction in the errors is very large. For instance, at low $Q^2$ (below
100\,GeV$^2$), the theoretical uncertainties in the ratio are fourth of
those in the three-jet distributions. The cited value of $\as(M_Z)$ as
determined from the measurements of $R_{3/2}$ is
$$
\as(M_Z) =
0.1179\pm0.0013\,\rm(stat.)
\,^{+0.0028}_{-0.0046}
\,\rm(exp.)
\,^{+0.0064}_{-0.0046}\,\rm(theo.)
\,.
$$
The dominant source of uncertainty is still the theoretical one which
calls for further efforts in improving the predictions by computing
even higher order corrections.
 
\section{Recent developments: predictions for multi-jet\\ event shapes}
\label{evshape}

There are two directions in computing higher-order corrections. One is
the exact fixed-order computations that I discussed previously by
considering the NLO corrections. Going beyond the NLO accuracy is very
difficult and so far has only been achieved for totally inclusive
quantities such as structure functions. For jet cross sections the
first step in order to make advances in this direction is the computation 
of inclusive jet and dijet cross sections at the next-to-next-to-leading
order (NNLO) accuracy. The recent advances in computing the NNLO
corrections in the crossed channel of electron-positron annihilation into
three jets \cite{3jetNNLO} raises hopes that for jet cross sections in
DIS the NNLO prediction will also be available in the not too distant
future.\footnote{Note that the naive iterative extension of the dipole
subtraction scheme to NNLO is not possible \cite{IRlimit0}.}  In order
to compute NNLO corrections to the multi-jet cross sections, a major
bottleneck is the computation of the necessary virtual corrections, and
I do not expect quick progress in this direction.  

The other possibility to improve the predictions is to resum the most
important logarithmic corrections, due to collinear and soft radiation,
to all orders, which leads to predictions at the next-to-leading
logarithmic (NLL) accuracy. Such computations are not available for jet
rates.  However, much progress has been achieved recently in resumming
the LL and NLL contributions to multi-jet event-shape distributions
\cite{Dasgupta:2003iq}.  These works lead to much deeper insight about the
structure of QCD cross sections. In particular, prior to these studies
it was believed that distributions of DIS observables, measured in the
current hemisphere in the Breit frame, were trivially related to their
well studied counterparts in electron-positron annihilation, where the
resummed logarithms are due to soft radiation over the whole phase
space (hence they are called global observables).  However, it was
found that there were also important single-logarithmic non-global
effects due to radiation into one hemisphere \cite{Dasgupta:2001sh}.
In this talk I want only to collect the currently available theoretical
information on multi-jet event shapes without going into the details of
the theoretical studies.


There are two multi-jet event shapes computed to NLL accuracy so far.
One quantifies the out-of-plane QCD radiation (the sum of the
momentum components perpendicular to the event plane), called $\Kout$,
defined precisely in \Ref{Kout}. The other is the $y_3$ observable
that is defined to be the largest value of the jet resolution variable
$\ycut$ such that the event is clustered into three jets with
$\kT$-clustering \cite{kTclusDIS}.
The $\Kout$ distribution has been studied experimentally in
\Ref{Everett}, however, with different definition of the observable as
done in the resummation computation, therefore, conclusions cannot be
drawn from the results.

\begin{figure}[ht]
\centerline{
\epsfxsize=60mm \epsfbox{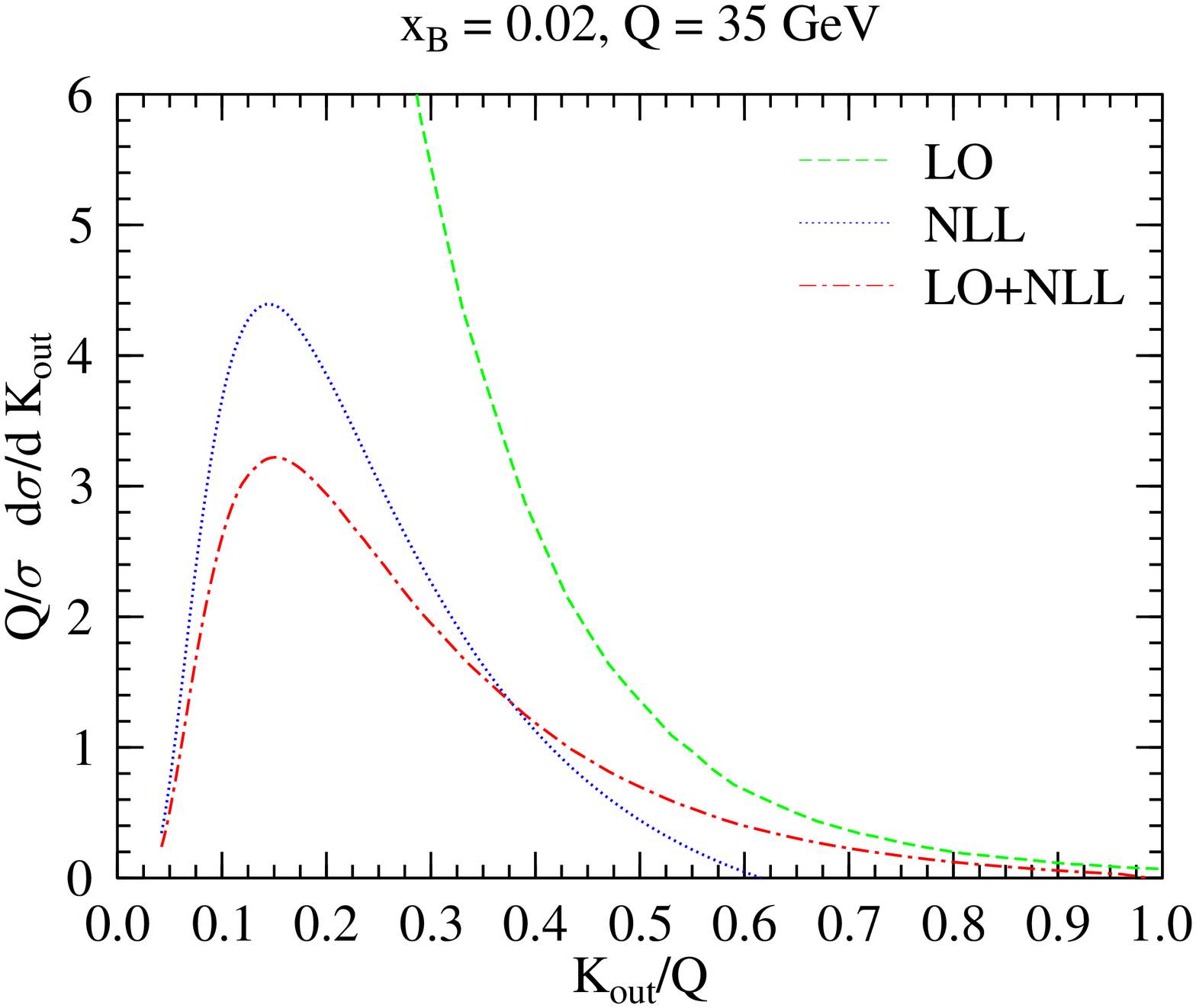}
\epsfxsize=60mm \epsfbox{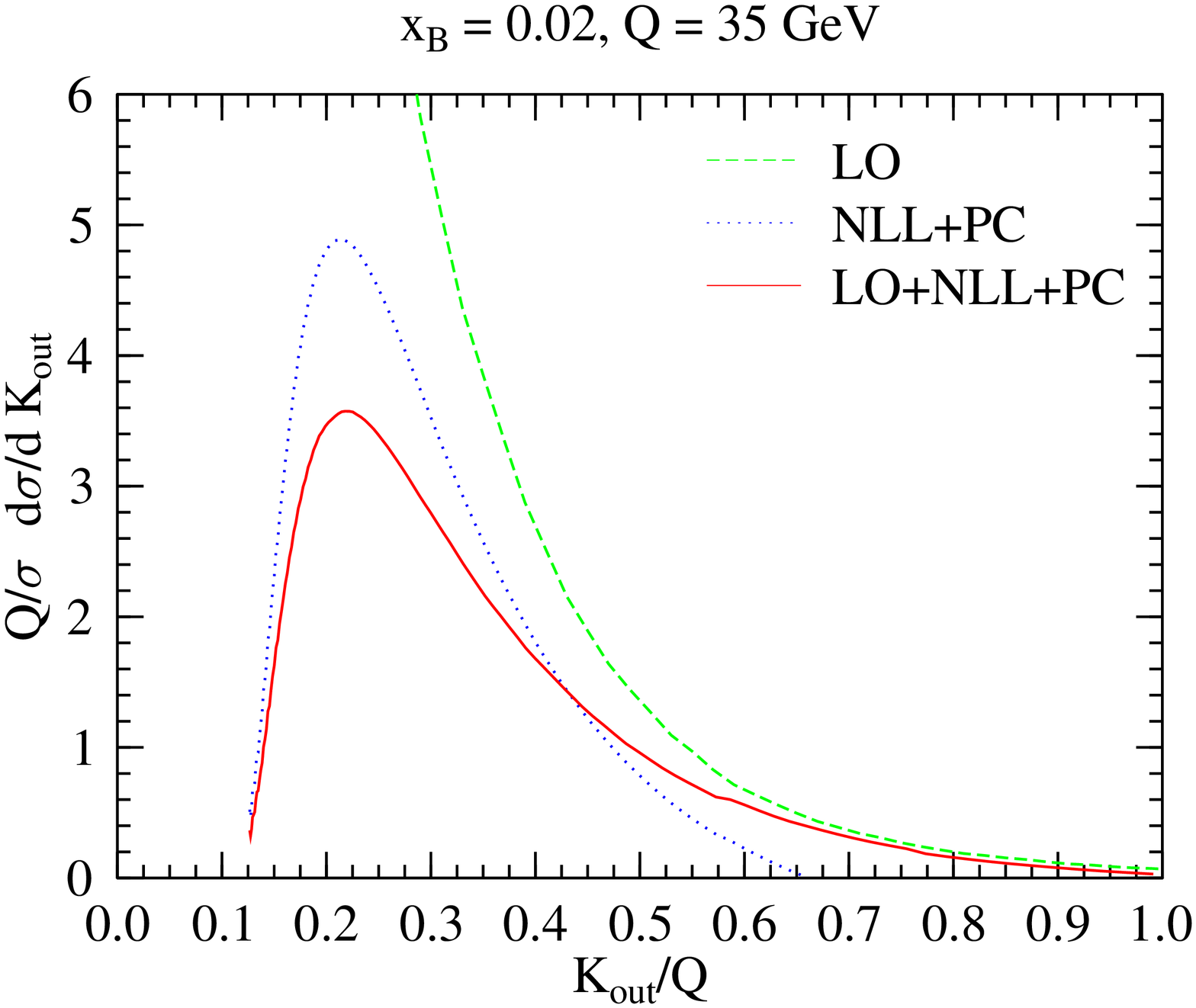}
}
\caption{The differential distribution of $\Kout$.
Left panel: perturbative computations. Right panel: perturative
predictions with power corrections.
\label{fig:Kout-LO+NLL}}
\end{figure}

One may ask why the computation of the NLO corrections is necessary if
resummed predictions are known. The reason is that the NLL and NLO
predictions are valid in rather distinct parts of the phase space, which
can be clearly seen on the left panel of \Fig{fig:Kout-LO+NLL}, where the
differential distributions in $\Kout/Q$ at fixed values of $Q^2 = (35
{\rm GeV})^2$ and $\xB =0.02$, normalized to the Born cross section
are presented.  The dotted line is the LO prediction, the dashed is the
NLL one. Expanding the NLL prediction in $\as$ and changing the leading
term to the exact LO one, we obtain the matched prediction shown with
the dash-dotted line. We see that in the $\Kout$-region where the best
precision experimental data can be collected, neither the fixed-order
nor the resummed values are reliable, but one should use the matched
prediction. On the right panel, I show the effect of including the power
corrections both to the NLL and the mathced predictions. The importance of
matching is clear also in this case.


In analysing the multihadron data collected in electron-positron
annihilation, very accurate theoretical description of event-shape
distributions was found with matched resummed and fixed-order predictions
improved with hadronisation corrections (see e.g. \cite{NTQCD98}). I
expect it will also be interesting to compare the HERA results for
multi-jet event-shape distributions to predictions of the same level.
Conclusions of such studies could also be important for analyses at the
LHC, where the presence of the incoming two hard partons in the event
means that even the dijet event shapes need at least four hard partons which
is a multi-jet event shape configuration in DIS. Thus dijet event
shapes at hadron colliders represent kinematical situations where NLL
resummations and power corrections are as yet untested. 

\begin{figure}[ht]
\centerline{
\epsfxsize=60mm \epsfbox{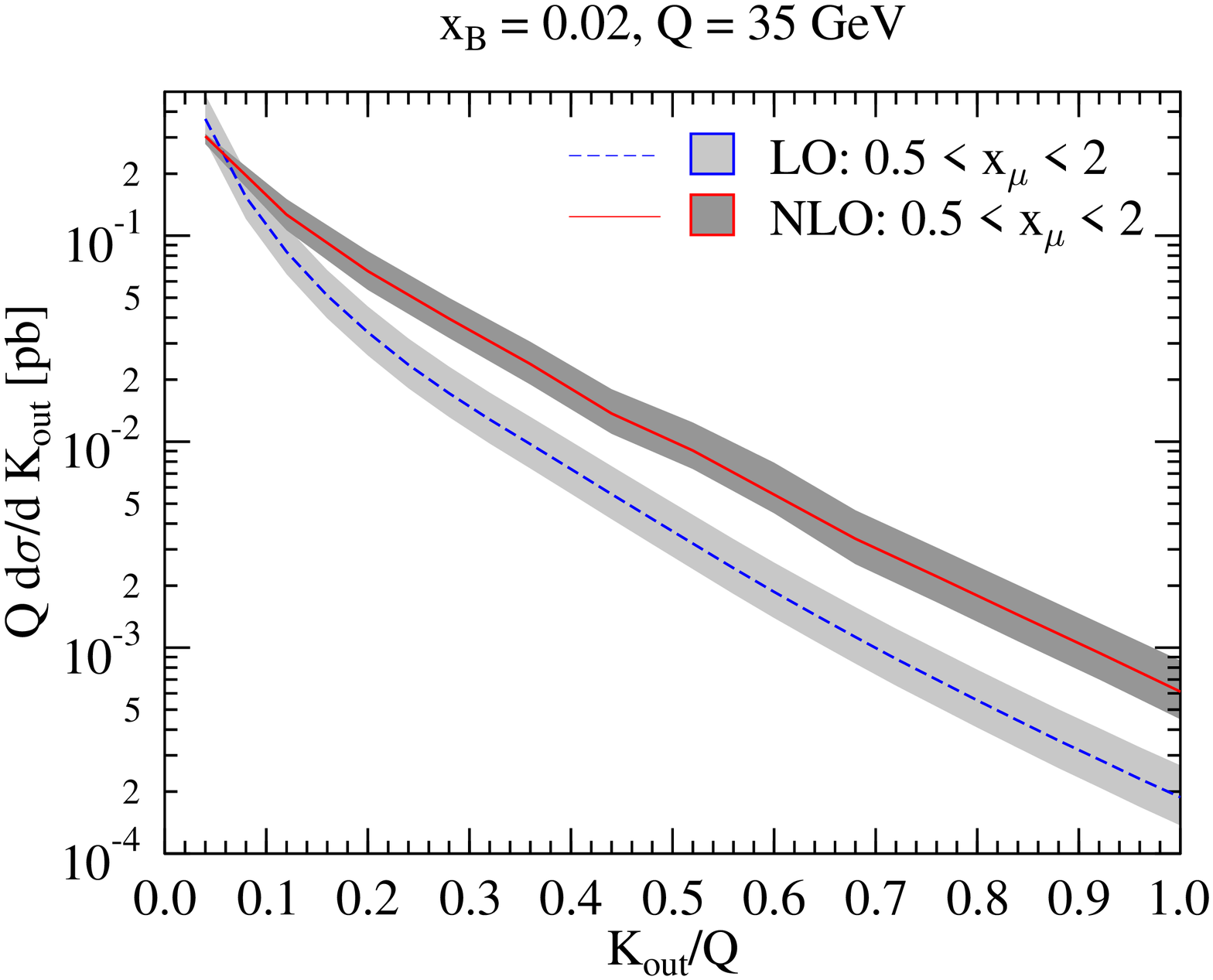}
\epsfxsize=60mm \epsfbox{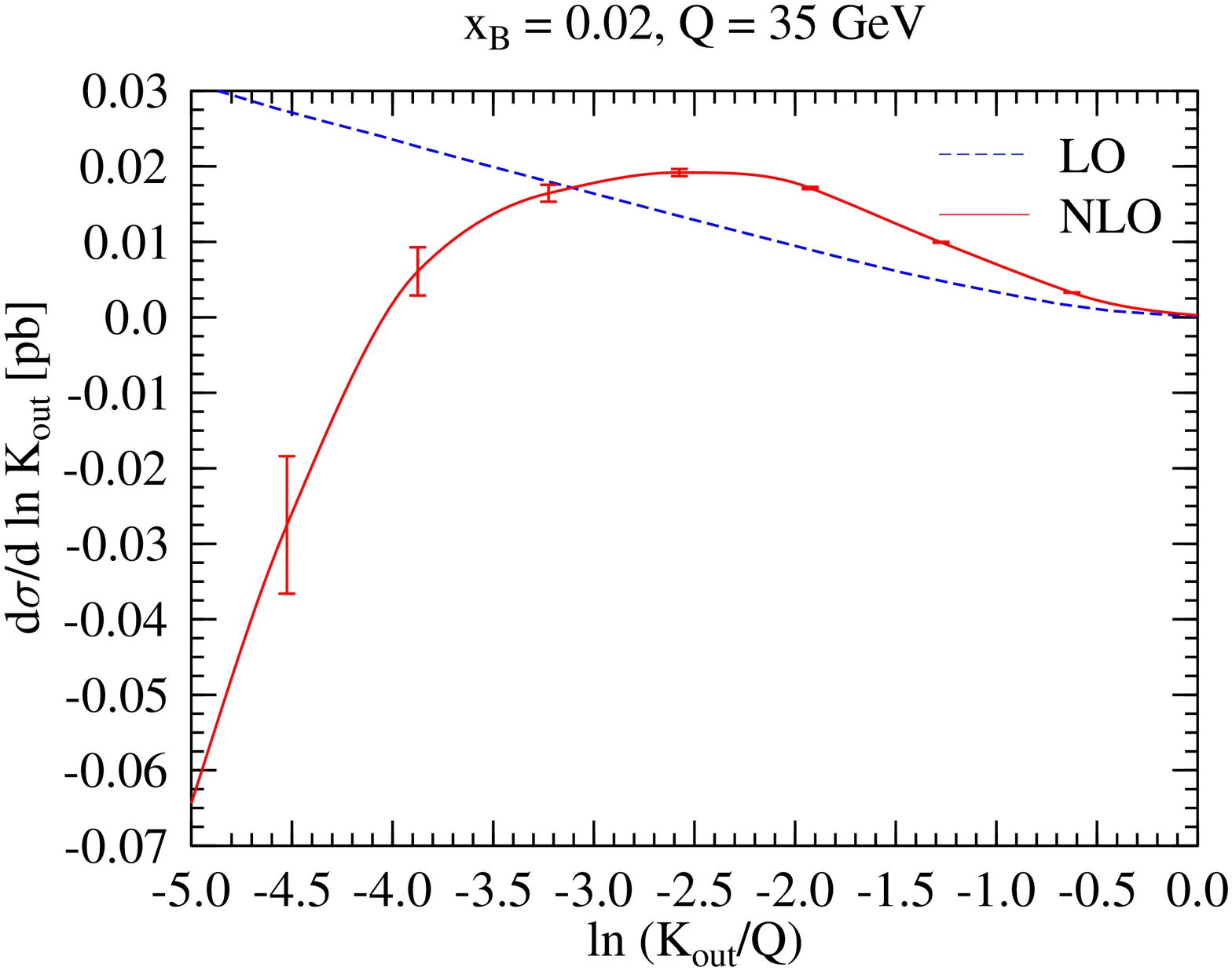}
}   
\centerline{
\epsfxsize=60mm \epsfbox{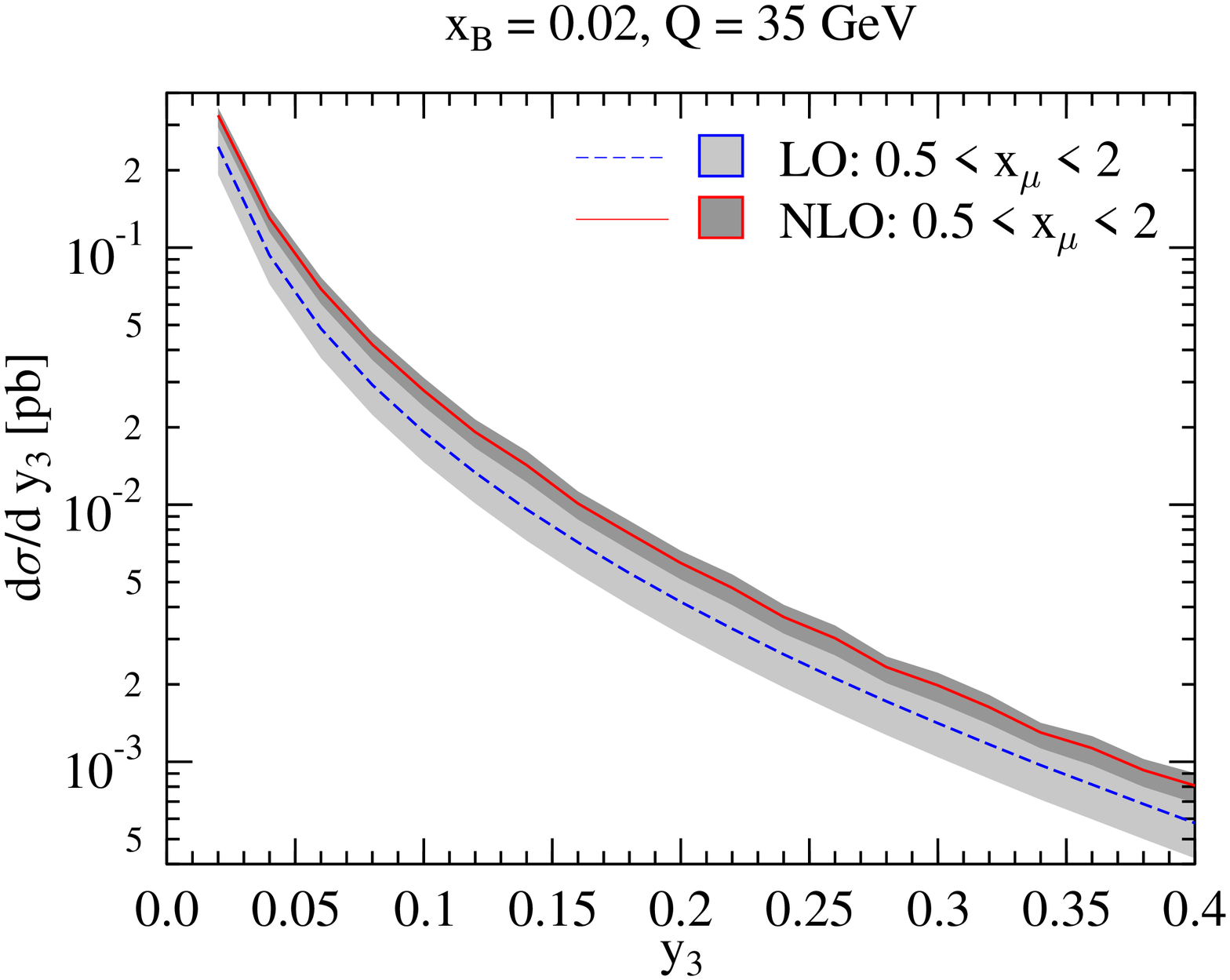}
\epsfxsize=60mm \epsfbox{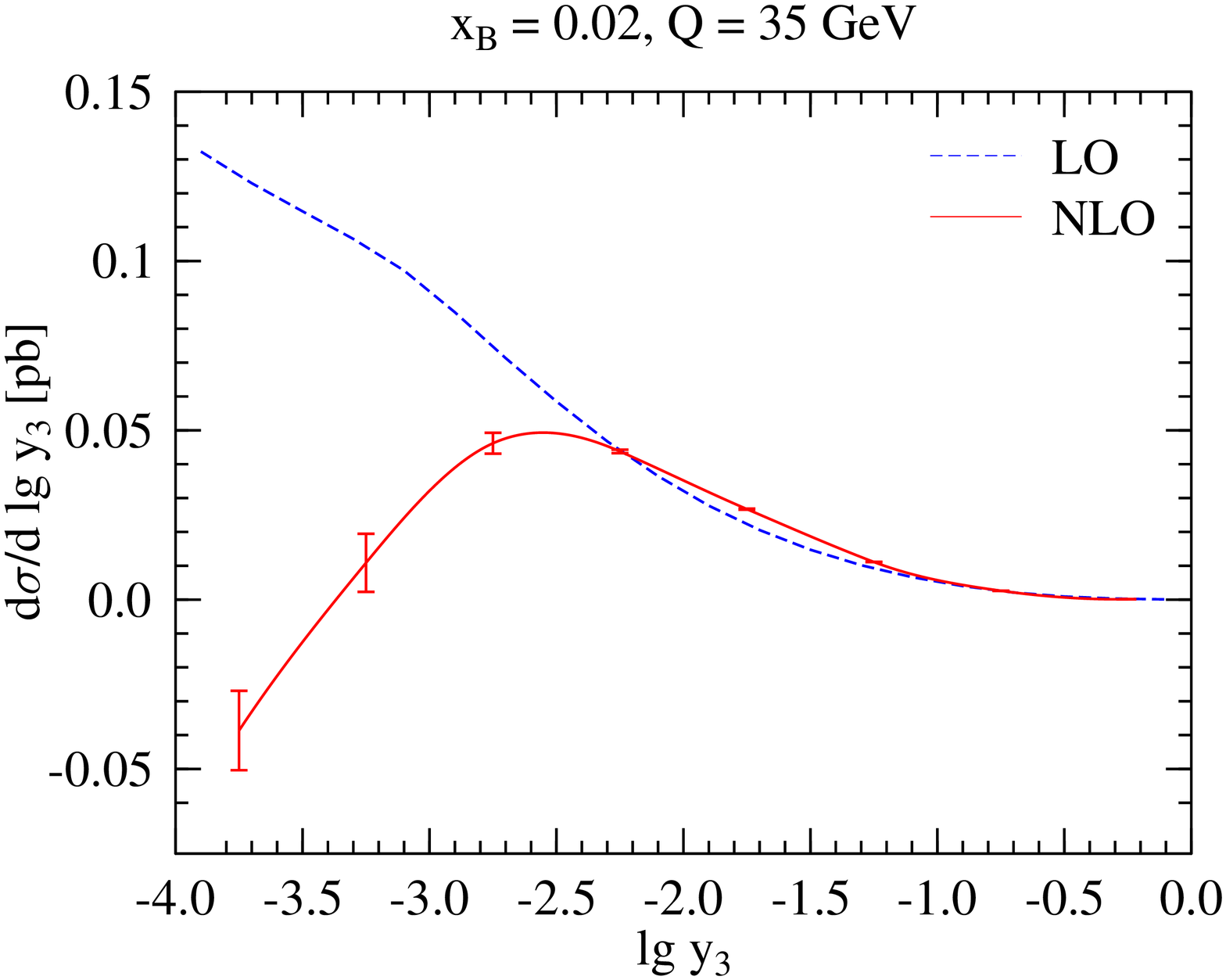}
}   
\caption{The differential distributions of the \Kout\ and $y_3$
observables at fixed value of $\xB$ and $Q^2$.  The left panel shows
the distributions as a function of $\Kout/Q$ and $y_3$, the right panel
shows the distributions as a function of $\ln \Kout/Q$ and $\lg y_3$ in
order to exhibit the logarithmic behaviour for small values of the
observable. Both the \LO\ predictions (dashed line) and the \NLO\
predictions (solid line) were obtained with the CTEQ6M \pdf s. The
errorbars indicate the uncertainty of the numerical integration that is
negligible for the computation at LO accuracy.
\label{fig:Kout}}
\end{figure}

The semi-automatic computation of NLL predictions with the program
{\sc caesar} is currently being interfaced to the output of the
{\sc nlojet++} code. With this interface matched fixed-order and
resummed predictions improved with power corrections will be obtained
in a semi-automatic way soon \cite{Banfiprivate}. Here I would like to
present NLO predictions for the distribution of the event-shape
observable \Kout, computed recently \cite{NTdisevshape}.  I used the
same definitions of the observables and performed the computations at
fixed values of the DIS kinematic variables $Q^2 = (35 {\rm GeV})^2$,
$\xB = 0.02$ as in the resummed computations \cite{caesarpage}.
Figure~\ref{fig:Kout} shows the LO and NLO predictions. The shaded bands
in the left panel correspond to the range of scales
 $1/2 \le x_R = x_F \le 2$. We find that the radiative corrections are
in general large, thus the scale-dependence reduces only relatively to
the cross sections. They also increase with increasing value of $\Kout$
because the phase
space for events with large out-of-plane radiation with three partons in
the final state (at LO) is much smaller than that with four partons in
the final state (real corrections). The boundary of the phase space in
$\Kout$ is about 20\,\% larger for the NLO computation than at LO. The
cross sections decrease rapidly with increasing $\Kout$. The small
cross section for medium or large values of $\Kout$ leaves the small
$\Kout$-region for experimental analysis.

In the small $\Kout$-region, the logarithmic contributions of the type
$\ln \Kout/Q$ are dominant as can be seen on the plot in the right
panel. At LO, the logarithmic dominance starts at about $\ln\Kout/Q=-2$,
at NLO, it starts at about $\ln\Kout/Q=-4$. Below these values the
cross section is a linear function of $\ln \Kout/Q$ and the fixed-order
predictions diverge with $\Kout \to 0$ with alternating signs, which
makes the resummation of these large logarithmic contributions
mandatory.  Reliable theoretical predictions can be obtained by
matching the cross sections valid at the NLO and NLL accuracy. This
matching is obtained by expanding the NLL prediction in $\as$ and
changing the first two terms in that expansion with the exact values of
the NLO computation. Qualitatively similar conclusions can be drawn from
the distributions for the observable $y_3$ \cite{NTdisevshape}, but the
corrections are smaller.
 
\section{Conclusions and Outlook}

In this talk I discussed the present status of predicting distributions
of multi-jet cross sections in lepton-proton scattering. The only existing
program for computing the three-jet observables at \NLO\ accuracy in
DIS is the {\sc nlojet++} code. This program  is well-tested, but has the
slight disadvantage that the $Z$-boson exchange diagrams are not
included. The predictions for three-jet rates agree well with the data
collected at HERA, although the main source of uncertainty remains the
theoretical one, which calls for taking into account the higher order
corrections. 

The other option of taking into account higher orders is the matching
with resummed predictions valid at the NLL accuracy. Recent years yielded
a lot of progress in this area of research. The {\sc caesar} program can
be used for computing NLL predictions to multi-jet distributions in a
semi-automatic way. I showed the importance of matching the NLO and NLL
predictions. Both are available for certain three-jet event-shape
observables, like the out-of-plane momentum \Kout and the $y_3$ variable.
The matching of the NLO and NLL predictions is expected to be available
soon \cite{Banfiprivate}.

\section*{Acknowledgments}
I am greatful to the organizers of the Ringberg workshop for their
invitation as well as for providing a pleasant atmosphere during the
meeting. This work was supported by the Hungarian Scientific Research
Fund grant OTKA T-038240.

\end{document}